\documentclass[epj]{svjour}

\usepackage{amsmath}
\usepackage{amssymb}
\usepackage{graphicx}

\begin{document}


\title{Deconstructing $^1$S$_0$ nucleon-nucleon scattering}
\author{Michael C. Birse}
\institute{Theoretical Physics Group, School of Physics and Astronomy,\\
The University of Manchester, Manchester, M13 9PL, UK\\}

\date{\today}

\abstract{
A distorted-wave method is used to analyse nucleon-nucleon scattering
in the $^1S_0$ channel. Effects of one-pion exchange are removed from
the empirical phase shift to all orders by using a modified effective-range 
expansion. Two-pion exchange is then subtracted in the distorted-wave Born 
approximation, with matrix elements taken between scattering waves for the
one-pion exchange potential. The residual short-range interaction shows a very 
rapid energy dependence for kinetic energies above about 100~MeV, suggesting 
that the breakdown scale of the corresponding effective theory is only 
270~MeV. This may signal the need to include the $\Delta$ resonance as
an explicit degree of freedom in order to describe scattering at these 
energies. An alternative strategy of keeping the cutoff finite to reduce 
large, but finite, contributions from the long-range forces is also 
discussed.
\PACS{{13.75.Cs} {12.39.Fe} {11.10.Gh} {21.20.Cb}}}
\maketitle

\vspace{10pt}

\section{Introduction}

Much time and effort has been spent pursuing the goal of systematically 
understanding nuclear forces through the techniques of effective field 
theory (EFT).\footnote{For reviews, see Refs.~\cite{border,bvkrev,eprev}.}
Unlike the low-energy interactions among pions, the nucleon-nucleon (NN)
interaction is nonperturbative, at least in some channels, and 
this had led to ongoing debates about the correct power counting
to use and about the role of the cutoff or regulator in these 
theories.

An EFT is built out of fields corresponding to the appropriate low-energy 
degrees of freedom -- pions, nucleons and photons for nuclear physics. Its 
Lagrangian (or Hamiltonian) contains all possible terms consistent with the 
symmetries of the underlying dynamics. To have any predictive power, 
therefore, the theory must be systematically expandable in powers of ratios 
of low-energy scales, denoted generically by $Q$, to those of the underlying
physics. Denoting a generic high-energy scale by $\Lambda_0$, the expansion 
parameter is $Q/\Lambda_0$ and so the convergence of the theory is expected 
to break down when $Q$ is of the order of the smallest of these scales.
Since the $\Lambda_0$ determine the sizes of the coefficients of the operators 
in the effective Lagrangian, fitting those coefficients to observables gives a 
measure of the scale of the underlying physics. This means that an EFT should 
always provide an indication of its own radius of convergence.

For nuclear physics, the relevant low-energy scales include the momenta 
of the nucleons and the mass of the pion. The scales of the underlying theory, 
Quantum Chromodynamics (QCD), include the one associated with the hidden 
chiral symmetry, $4\pi F_\pi$, as well as the masses of the nucleons and 
$\rho$, $\omega$ and other heavy mesons. We might therefore hope that 
$\Lambda_0$ would be of the order of several hundred MeV and so we could 
construct an EFT valid for momenta up to about $2m_\pi$. However there could 
be other important scales in QCD and so we need to let the data determine 
the range of validity of our EFT.

The starting point for all applications of EFTs to nuclear forces is 
Weinberg's observation \cite{wein1,wein2} that nonrelativistic loop integrals 
are enhanced relative to those in relativistic theories, being of order 
$Q$ rather than $Q^2$ as in, for example, mesonic chiral perturbation 
theory (ChPT) \cite{ecker}. Naive dimensional analysis (``Weinberg power 
counting") indicates that the leading terms in the 
NN force, one-pion exchange (OPE) and a momentum-independent contact term, 
are of order $Q^0$. This would imply that these terms are still 
perturbative, each iteration leading to an extra power of $Q$ from the 
integral. The resulting theory would not be able to describe bound
states -- nuclei.

In order to generate low-energy bound or virtual states, we need to identify 
additional low-energy scales that can promote the leading interactions to 
order $Q^{-1}$. The first such scales to be noted were provided by the 
$S$-wave scattering lengths, which lead to EFTs where the leading contact terms 
must be iterated \cite{bvk,vk,ksw1,ksw2}. At very low energies, where pion-exchange 
forces can be replaced by contact interactions, this leads to a ``pionless" EFT, 
which provides a field-theoretic realisation of the effective-range expansion
\cite{bethe,bj}. This formulation allows that, much older, approach to be extended
to systems of three or more particles and to processes involving electromagnetic 
or weak interactions.

To describe physics on momentum scales of the order of $m_\pi$ or larger, 
we need to keep pion-exchange forces explicitly. As already mentioned, OPE is 
of order $Q^0$ in naive dimensional analysis, implying that it should be
treated  perturbatively. Kaplan, Savage and Wise \cite{ksw1,ksw2} used this to set 
up an EFT based on a power counting (``KSW counting") in which the momenta, 
$m_\pi$ and the inverse scattering lengths were the low-energy scales. 
However it was soon shown that the resulting expansion fails to converge in 
the $^3S_1$ wave \cite{ch1,ch2,fms,bbsvk}. This suggests that there is another 
low-energy scale related to the strength of OPE that would justify iterating
that potential as well the leading contact interaction. 
The perturbative treatment of OPE did seem to be valid in the $^1S_0$ 
channel, but Fleming, Mehen and Stewart \cite{fms} found large changes to 
the strengths of lower-order contact interactions when higher-order terms 
were included. They concluded that the expansion parameter of the theory 
might be as large as $\sim1/2$.

By dividing a factor of $1/M_{\scriptscriptstyle N}$ out of the Hamiltonian,
we can express the strength of the OPE potential in terms of a momentum 
scale,
\begin{equation}
\lambda_{\pi{\scriptscriptstyle NN}}
=\frac{16\pi F_\pi^2}{g_{\scriptscriptstyle A}^2
M_{\scriptscriptstyle N}}\simeq 290\;\mbox{MeV},
\label{eq:lambdapi}
\end{equation}
where, to lowest order in the chiral expansion, the Gold\-berger-Treiman relation has been used to express the $\pi$N coupling constant in terms of 
$g_{\scriptscriptstyle A}$. Note that $\lambda_{\pi{\scriptscriptstyle NN}}$ 
is composed of high-energy scales in chiral perturbation theory, $4\pi F_\pi$ 
and $M_{\scriptscriptstyle N}$. Treating it as a high-energy scale leads 
to the perturbative treatment of OPE and KSW counting for the associated 
short-range interactions. However the numerical value of 
$\lambda_{\pi{\scriptscriptstyle NN}}$ is only about twice $m_\pi$, which 
may explain some of the problems with KSW counting found in 
Refs.~\cite{ch1,ch2,fms,bbsvk}. If instead we identify 
$\lambda_{\pi{\scriptscriptstyle NN}}$ as a low-energy scale, then the
OPE potential is promoted to to order $Q^{-1}$, implying that it
should be iterated. 

In this paper, I examine the consequences of treating nonperturbatively 
both OPE and the leading, energy-independent contact interaction in the 
$^1S_0$ channel. Iterating OPE generates to set of distorted waves (DWs). 
Further scattering between these waves can be described by a residual
$K$ matrix that contains the effects of short-range interactions
and other long-range-forces, such as two-pion exchange (TPE).
The short-range interaction is strong in this channel and so this
$K$ matrix can be expanded using a DW or ``modified" effective-range
expansion \cite{bethe,vhk,bkps,sf}. This in contrast to the peripheral
channels studied in Refs.~\cite{bmcg,bir07} where the scattering is 
weak and a distorted-wave Born approximation (DWBA) can be used to
determine the effective potentials from empirical phase shifts.

This approach is based on an EFT whose low-energy scales are the
nucleon momenta, the inverse of the $^1S_0$ scattering length, 
$m_\pi$ and $\lambda_{\pi{\scriptscriptstyle NN}}$. The 
power counting for it can be found by analysing the scale 
dependence of the short-range potential with the help of the
renormalisation group (RG) \cite{wrg}. For present purposes, the RG 
methods developed in Refs.~\cite{bmr,bb1,bb2,bir05} for nonrelativistic 
scattering show that the counting follows from the behaviour of the 
wave functions near the origin which, in turn, is controlled by the 
singularity of the long-range potential as $r\rightarrow 0$. 
The spin-singlet channels see only the central
part of OPE, which has the usual Yukawa form, behaving like  $1/r$
near the origin. Its distorted waves have the same power-law behaviours as 
free waves and hence iteration of this potential does not alter the
power counting for the terms of the short-range interaction 
\cite{bb1}. This is quite different from the situation in the spin-triplet 
channels where the $1/r^3$ singularity of the OPE tensor potential leads to 
a counting in which the leading contact interactions are substantially 
promoted \cite{ntvk,bir05,bir07}.

Unlike the peripheral singlet waves, the $^1S_0$ channel has a virtual 
state at very low energy and so the scattering is strong near threshold. 
The appropriate power counting corresponds to an expansion around a 
nontrivial fixed point of the RG and is similar to the effective-range 
expansion \cite{bethe} for pure short-range interactions. The one notable 
difference in the presence of OPE is a leading-order contact term 
proportional to $m_\pi^2/\lambda_{\pi{\scriptscriptstyle NN}}$. This is 
need\-ed to renormalise a logarithmic divergence, as found first by Kaplan 
\textit{et al.}~\cite{ksw1,ksw2} in the context of a perturbative treatment of OPE. 
Although OPE is iterated to all orders here, the terms of the short-range 
potential in this channel can still be organised according to a modified 
version of KSW counting that takes into account the extra scale 
$\lambda_{\pi{\scriptscriptstyle NN}}$.\footnote{In practice, terms in the 
present approach that contain powers of the scale 
$\lambda_{\pi{\scriptscriptstyle NN}}$ cannot be 
disentangled from lower-order contributions with the same energy
dependences. The present analysis of the $^1S_0$ channel could therefore
be viewed as an application of the original KSW counting, the iteration of 
OPE being purely for computational convenience, to avoid the need for 
evaluating terms up to fourth order in perturbation theory.} 

The residual short-range potential of this theory can be determined 
directly from empirical phase shifts, via a DW effective-range expansion 
which removes the effects of iterated OPE \cite{bb1}. This expansion
has previously been applied to $^1S_0$ NN scattering by Steele and
Furnstahl \cite{sf}, who noted the appearance of scales of the order
of $2m_\pi$ in the resulting short-distance coefficients and suggested 
that removal of two-pion exchange (TPE) might improve the radius of 
convergence. As a coordinate-space approach, it has close connections to 
ones developed by other groups \cite{pvra1,pvra2,pvra3,spm,pv09}, where interaction 
strengths are also related to logarithmic derivatives of the wave 
functions at small radii. 

Previous applications of this method to the $^1S_0$ channel \cite{sf}
and of a related DWBA approach to peripheral partial waves 
\cite{bmcg,bir07} have found significant energy dependences in the 
residual interactions after removal of the effects of OPE. These suggest 
that other long-range forces are also important. Interactions that 
have been well studied in ChPT are TPE at orders $Q^2$ and $Q^3$ \cite{kbw} 
(see also Ref.~\cite{nij99}), as well as the order-$Q^2$ recoil correction 
to OPE \cite{friar} and the leading $\pi\gamma$-exchange force \cite{fvkpc}. 
As terms of order $Q^2$ or higher, all these should be treated as 
perturbations in the power counting used here. This means that their
effects can be removed using the DWBA, evaluating their matrix elements 
between DWs of OPE and subtracting them from the residual $K$ matrix, as
done in Refs.~\cite{bmcg,bir07}.

If we were to iterate these higher-order terms by solving the 
Schr\"od\-inger equation with the full potential, then their singularities 
would alter the forms of the short-dis\-tance wave functions. In general, 
this destroys any consistent power counting, something that has been 
observed many times, as in Ref.~\cite{pbc}, for example. Such problems can 
be avoided if we work with a finite cutoff, keeping it within the domain 
of validity of the EFT as discussed in Refs.~\cite{eg,bir09}.
The price to be paid for this is that the resulting
effective potential will contain artefacts of the cutoff, that is, pieces 
that would vanish if one were able to take the regulator scale to infinity.
As a result, the form of the short-distance potential is not determined solely
by the physical scales of the system -- the finite cutoff also plays a crucial 
role. In the context of standard few- and many-body techniques, which are
designed to solve the Schr\"odinger equation with a given potential, this 
price may be unavoidable. 
 
On the other hand, if we treat the higher-order terms in the potential 
perturbatively, as dictated by the pertinent power counting, then these 
problems do not arise. To any given order, the EFT contains the necessary 
counterterms to cancel any divergences. Having done that, we can 
then make all cutoff artefacts small by taking the regulator to be as large 
as we like. The only proviso is that, before doing so, we need to have 
determined the counting that applies to our system \cite{bir09}. 
The first application of this method, to $L\geq 2$ spin-singlet partial 
waves, did not require any renormalisation \cite{bmcg}. However 
the higher-order terms of the chiral TPE potential are highly singular as 
$r\rightarrow 0$ and so, in all other waves, their matrix elements are 
divergent at small radii. For example, the $^1P_1$ wave requires a single 
order-$Q^2$ term to cancel an energy-independent divergence \cite{ihb},
and two terms are needed in each spin-triplet wave \cite{bir07,pv09},
as expected from the power counting in the presence of tensor OPE
\cite{ntvk,bir05}. Once we have renormalised these matrix elements
of the higher-order potentials, they can be treated in perturbation 
theory. 

Here, I  use this method to ``deconstruct" scattering in the $^1S_0$ 
channel. First, the effects of iterated OPE are removed from the empirical 
phase shifts, and then the DWBA matrix elements of the order-$Q^{2,3}$ 
chiral potentials are subtracted to leave a residual short-range 
interaction. In this channel, the presence of a low-energy virtual state 
means that the leading, energy-independent contact interaction must be 
treated to all orders. Doing so introduces admixtures of irregular wave 
functions into the solutions of the Schr\"odinger equation. These increase 
the degrees of divergence of the DWBA matrix elements of TPE. Nonetheless, 
KSW power counting (the counting for perturbations around the nontrivial 
fixed point) provides the necessary counterterms to renormalise these 
divergences. As a result, the short-range interaction tends to a form that 
is independent of regulator, provided that regulator is chosen to be high 
enough in momentum or, in coordinate space, small enough in radius.

In a closely related approach, Shukla \textit{et al.}~\cite{spm} extracted 
a boundary-condition parameter from the logarithmic derivatives of the $^1S_0$ 
wave functions near the origin, and expressed this in the form of a
short-range interaction strength. However they did not strictly follow the 
power counting advocated here, but either iterated the order-$Q^3$ potential 
to all orders by solving the Schr\"odinger equation, or treated the whole 
long-range potential to second order. The singularities of the TPE potential 
mean that, for the reasons mentioned above, these calculations were restricted 
to large cutoff radii, $\gtrsim 1$~fm. More recently, Pav\'on Valderrama 
\cite{pv09} has also applied the DW approach to the $^1S_0$ and 
$^3S_1$--$^3D_1$ channels, within the framework of the power countings 
that follow from iterating OPE \cite{bb1,bir05}. In that work, he fitted a 
polynomial form for the short-range interaction to the empirical phase shifts, rather than extracting such a potential directly from them, as done in 
Ref.~\cite{spm} and here. 

Contrary to the hopes of Steele and Furnstahl \cite{sf} and to the 
situation in the peripheral waves \cite{bmcg,bir07}, but consistent with 
the observations of Pav\'on Valderrama \cite{pv09}, the current approach 
leads to short-range interactions that show an even stronger energy 
dependence after subtraction of the order-$Q^{2,3}$ potentials. This 
suggests that the breakdown scale of the resulting EFT may be only of the 
order of $2m_\pi$. This is a serious problem for hopes of applying the EFT 
discussed here to larger nuclear systems, where typical momentum scales 
are of that size. One possibility is to look for an EFT in which
at least some parts of the TPE potential are promoted to order $Q^{-1}$ 
and so should be iterated. However that would require identifying further 
low-energy scales, by analogy with $\lambda_{\pi{\scriptscriptstyle NN}}$.

An alternative way round this might be to work with a finite regulator, 
chosen to optimise the convergence of the EFT \cite{spkc}. For example,
Pav\'on Valderrama \cite{pv09}, like the authors of Ref.~\cite{spm}, used  
relatively large values for the cutoff radius, $\gtrsim 0.6$~fm. This was 
not essential since the necessary counterterms are present to cancel 
perturbatively all divergences of the TPE matrix elements, but working with 
cutoffs in this range does lead to results with a greater apparent radius 
of convergence. Regulator scales corresponding to even larger radii are 
commonly used in the ``Weinberg scheme" \cite{em,egm,eg}, where the full 
chiral potential is iterated to all orders. However, as I shall discuss, 
they would play a quite different role in the present context, taming large 
but not divergent contributions from the long-range potentials. Similar 
applications of finite regulators as tools to improve the convergence of 
EFTs have been suggested by various authors \cite{spkc,dsgs,dgss}, most recently 
by Beane \textit{et al.}~\cite{bkv} for calculations of S-wave NN 
scattering that treat OPE perturbatively.

This paper is organised as follows. First, in Sec.~2, I outline the DW 
method used to remove the effects of OPE and hence extract an effective 
short-range interaction directly from the phase shift. Then, in Sec. 3, 
I use the DWBA to subtract perturbatively the effects of the order-$Q^{2,3}$ 
chiral potentials. This leaves a residual short-range interaction whose only ``contamination" from long-range forces starts at order $Q^4$. Finally, 
in Sec.~4, I discuss the implications of the strong energy dependence 
displayed by this potential and comment on the approaches suggested in 
Refs.~\cite{bkv,spm}.

\section{One-pion exchange}

Having identified $\lambda_{\pi{\scriptscriptstyle NN}}$ as a low-energy scale 
in NN scattering, we need to iterate OPE to all orders. In a coor\-dinate-space
approach, this is done by solving the Schr\"od\-inger equation in differential
form. For the $^1S_0$ channel, this is
\begin{equation}
\left[-\,\frac{1}{M_{\scriptscriptstyle N}}\left(\frac{{\rm d}^2}{{\rm d}r^2}
+\frac{2}{r}\,\frac{{\rm d}}{{\rm d}r}\right)
+V_{\rm OPE}(r)\right]\psi(p;r)
=\frac{p^2}{M_{\scriptscriptstyle N}}\,\psi(p;r).
\label{eq:se}
\end{equation}
Here $p$ is the on-shell relative momentum which is related to the lab kinetic 
energy, $T$, by $T=2p^2/M_{\scriptscriptstyle N}$. The central piece of the 
lowest-order OPE potential is
\begin{equation}
V_{\rm OPE}(r)=-f_{\pi{\scriptscriptstyle NN}}^2
\frac{e^{-m_\pi r}}{r},
\end{equation}
with $f_{\pi{\scriptscriptstyle NN}}$ denoting the pseudovector $\pi$N coupling 
constant which, in the chiral limit, is given by
\begin{equation}
f_{\pi{\scriptscriptstyle NN}}^2=\frac{g_{\scriptscriptstyle A}^2 m_\pi^2}
{16\pi F_\pi^2}.
\end{equation}

If the short-range interactions are strong enough to generate a low-energy bound
or virtual state, then we need both regular and irregular solutions to this 
equation. Near the origin, these have the expansions
\begin{eqnarray}
\psi_R(p;r)&=& A_R(p)\bigl(1+{\cal O}(r)\bigr),\cr
\noalign{\vspace{5pt}}
\psi_I(p;r)&=& A_I(p)\left(\frac{1}{r}-M_{\scriptscriptstyle N}
f_{\pi{\scriptscriptstyle NN}}^2\ln(\lambda r)+{\cal O}(r^2)\right),
\label{eq:dwRIexp}
\end{eqnarray}
where $A_{R,I}$ are overall normalisation constants. It is convenient to 
choose these so that as $r\rightarrow\infty$ the functions have the 
asymptotic forms,
\begin{eqnarray}
\psi_R(p;r)&\rightarrow& \frac{1}{pr}\sin\bigl(pr+\delta_R(p)\bigr),\cr
\noalign{\vspace{5pt}}
\psi_I(p;r)&\rightarrow& \frac{1}{pr}\sin\bigl(pr+\delta_I(p)\bigr),
\label{eq:dwRIasy}
\end{eqnarray}
where $\delta_R(p)$ is the usual phase shift induced 
by OPE alone. It is also convenient to introduce a second irregular solution
which is asymptotically out of phase with the regular one by $\pi/2$,
\begin{equation}
\psi_{I2}(p;r)\rightarrow \frac{1}{pr}\cos\bigl(pr+\delta_R(p)\bigr).
\end{equation}
This is given by the linear superposition 
\begin{equation}
\psi_{I2}(p;r)=\frac{\cos\bigl(\delta_R(p)-\delta_I(p)\bigr)\,
\psi_R(p;r)-\psi_I(p;r)}
{\sin\bigl(\delta_R(p)-\delta_I(p)\bigr)}.
\label{eq:dwI2}
\end{equation}

Linear superposition can then be used to construct a 
wave function with the observed phase shift $\delta(p)$:
\begin{equation}
\psi_>(p;r)=\cos\tilde\delta(p)\,\psi_R(p;r)
+\sin\tilde\delta(p)\,\psi_{I2}(p;r),
\label{eq:dwext}
\end{equation}
where $\tilde\delta(p)=\delta(p)-\delta_R(p)$ is the additional shift 
caused by shorter-range forces than OPE. This wave function is the same 
as one constructed by Ruiz Arriola and Pav\'on Valderrama using the 
variable-phase method \cite{pvra1,pvra2,pvra3}. Having it in a form that can 
be decomposed into regular and irregular pieces will be essential for the 
approach used here.

If these short-range forces are not resolved at the energies of interest, they
may be represented by any convenient choice of regularised contact interaction.
For example, Shukla \textit{et al.}~\cite{spm} use a square well with a sloping
floor. Here I take a $\delta$-shell form,
\begin{equation}
V(p;r)=\frac{1}{4\pi R_0^2}\,\widetilde V_S(p)\,\delta(r-R_0),
\end{equation}
like that previously used in Refs.~\cite{bmcg,bir07}. In a standard EFT 
treatment, the strength $\widetilde V_S(p)$ of the short-range potential is 
taken to be a low-order polynomial in $p^2$, whose coefficients are then fitted 
to the empirical phase shifts. Here, I make no assumptions about the 
form of $\widetilde V_S(p)$ but instead extracted it directly from the phase 
shifts. This makes it possible to examine where an expansion in powers of $p^2$ 
might be valid, and where it breaks down.

Outside the shell $r=R_0$, the wave function has the form $\psi_>(p;r)$,
containing the observed phased shift. For $r<R_0$, I take it to be
the solution of Eq.~(\ref{eq:se}) which is regular at the origin, 
$\psi_R(p;r)$ (in contrast to, for example, Ref.~\cite{spm} where only the 
short-range potential is present for $r<R_0$). For a solution to the full
Schr\"odinger equation, these pieces must match at $r=R_0$, 
$\psi_>(R_0)=\psi_R(R_0)$. The discontinuity in their radial derivatives is 
then related to the strength of the $\delta$-shell potential. We can thus
use this discontinuity to determine the strength $\widetilde V_S(p)$ needed
to reproduce the observed phase shifts:
\begin{equation}
\widetilde V^{(2)}_S(p)=\frac{4\pi R_0^2}{M_{\scriptscriptstyle N}}\,
\frac{\psi_>^\prime(p;R_0)-\psi^\prime_R(p;R_0)}{\psi_>(p;R_0)}.
\label{eq:VS2}
\end{equation}

This shows that the observed phase shift and the short-range effective
potential are connected through the logarithmic derivative of the asymptotic 
wave function, as has previously been noted in other 
coordinate-space approaches \cite{spm,pv09,pvra1,pvra2,pvra3}. The superscript (2) here 
signals the fact that long-range forces of order $Q^2$ and higher have not 
been subtracted and so their effects are subsumed within this short-range
potential.

Since the $^1S_0$ channel has a low-energy virtual state, this potential
lies close to a nontrivial fixed point. As noted in Refs.~\cite{bb1,bb2},
it is therefore more convenient to look at the behaviour of 
$1/\widetilde V^{(2)}_S(p)$. This can be written
\begin{equation}
\frac{1}{\widetilde V^{(2)}_S(p)}=\frac{M_{\scriptscriptstyle N}}{4\pi}\,
\frac{\psi_>(p;R_0)\,\psi_R(p;R_0)}{W_>(p)},
\label{eq:invVS2}
\end{equation}
where $W_>(p)$ is the Wronskian
\begin{equation}
W_>(p)=r^2\Bigl(\psi_>^\prime(p;r)\,\psi_R(p;r)
-\psi^\prime_R(p;r)\,\psi_>(p;r)\Bigr).
\end{equation}
In terms of the regular and irregular solutions $\psi_{R,I}$, this becomes
\begin{eqnarray}
\frac{1}{\widetilde V^{(2)}_S(p)}
&=&-\,\frac{M_{\scriptscriptstyle N}}{4\pi W_I}
\Bigl(\!\cot\tilde\delta(p)\sin\bigl(\delta_R(p)-\delta_I(p)\bigr)
\psi_R(p;R_0)^2\cr
&&\qquad\qquad\;+\cos\bigl(\delta_R(p)-\delta_I(p)\bigr)\,
\psi_R(p;R_0)^2\cr
\noalign{\vspace{5pt}}
&&\qquad\qquad\;-\psi_R(p;R_0)\,\psi_I(p;R_0)\Bigr),
\end{eqnarray}
where
\begin{equation}
W_I(p)=r^2\Bigl(\psi_I^\prime(p;r)\,\psi_R(p;r)
-\psi^\prime_R(p;r)\,\psi_I(p;r)\Bigr).
\end{equation}

This Wronskian $W_I$ is, of course, independent of $r$ but two useful 
expressions can be found by evaluating it at very large and very small $r$. 
Inserting the asymptotic forms of the solutions as $r\rightarrow\infty$, 
Eq.~(\ref{eq:dwRIasy}), gives
\begin{equation}
W_I(p)=\frac{\sin\bigl(\delta_R(p)-\delta_I(p)\bigr)}{p},
\end{equation}
while using the expansions around $r=0$, Eq.~(\ref{eq:dwRIexp}), gives
\begin{equation}
W_I=-A_R(p) A_I(p).
\end{equation}
With these two expressions we can rewrite Eq.~(\ref{eq:invVS2}) as
\begin{eqnarray}
\frac{1}{\widetilde V^{(2)}_S(p)}&=&-\,\frac{M_{\scriptscriptstyle N}}{4\pi}
\biggl(p\cot\tilde\delta(p)\,\psi_R(p;R_0)^2\cr
&&\qquad\qquad+p\cot\bigl(\delta_R(p)-\delta_I(p)\bigr)\,\psi_R(p;R_0)^2\cr
\noalign{\vspace{5pt}}
&&\qquad\qquad+\frac{\psi_R(p;R_0)\,\psi_I(p;R_0)}{A_R(p) A_I(p)}\biggr).
\label{eq:invVS2I}
\end{eqnarray}
At this point it is useful to introduce the $K$ matrix describing the 
scattering between the DWs of the OPE potential \cite{bb1},
\begin{equation}
\widetilde K(p)=-\,\frac{4\pi}{M_{\scriptscriptstyle N}p}\,
\tan\tilde\delta(p),
\label{eq:DWK}
\end{equation}
where, in the numerical results below, $\tilde\delta(p)$ is the difference 
between the empirical phase-shift determined by the Nijmegen group
\cite{nij93} and that generated by OPE alone.
This allows us to express Eq.~(\ref{eq:invVS2I}) in the form
\begin{eqnarray}
\frac{1}{\widetilde V^{(2)}_S(p)}&=&\frac{\psi_R(p;R_0)^2}{\widetilde K(p)}\cr
\noalign{\vspace{5pt}}
&&-\,\frac{M_{\scriptscriptstyle N}}{4\pi}\,
\biggl(p\cot\bigl(\delta_R(p)-\delta_I(p)\bigr)\,\psi_R(p;R_0)^2\cr
\noalign{\vspace{5pt}}
&&\qquad\qquad+\frac{\psi_R(p;R_0)\,\psi_I(p;R_0)}{A_R(p) A_I(p)}\biggr).
\label{eq:invVSKS}
\end{eqnarray}

Comparing Eq.~(\ref{eq:invVSKS}) with the analogous expressions in 
Ref.~\cite{bb1}, we see that the first term is (the inverse of) the 
short-range $K$-matrix divided by the square of the DW $\psi_>(p;r)$ at 
$r=R_0$. The remaining two terms arise from the loop integral
over DWs. The first of these can depend nonanalytically on low-energy scales
through the phase shifts $\delta_{R,I}$. It removes nonanalytic dependences
due to OPE from the empirical $K$ matrix to leave a short-range potential
that can be expanded in powers of the low-energy scales, for example 
with a DW version of the effective-range expansion, as described in
Ref.~\cite{bb1}. 

The last term can be expressed in terms of the power series expansions of 
$\psi_{R,I}$ around the origin. Since the Wronskian cancels the overall
normalisation factors that can contain nonanalytic behaviour, this term 
is analytic in the low-energy scales. The presence of the irregular
solution means that it diverges as the radial cutoff $R_0\rightarrow 0$. Using 
the expansions (\ref{eq:dwRIexp}), we find that the inverse of the potential
behaves for small $R_0$ as
\begin{equation}
\frac{1}{\widetilde V^{(2)}_S(p)}=-\,\frac{M_{\scriptscriptstyle N}}{4\pi}
\left(\frac{1}{R_0}+\frac{m_\pi^2}{\lambda_{\pi{\scriptscriptstyle NN}}}
\ln(\mu R_0)+{\cal O}(R_0^0)\right).
\label{eq:VS2div}
\end{equation}
If we define a corresponding momentum cutoff scale, $\Lambda\propto 1/R_0$, 
then we can see that the first term in this expansion is just the familiar linear
divergence of the NN loop diagram. The second term is the logarithmic divergence
proportional to $m_\pi^2$, whose counterterm was identified by Kaplan 
\textit{et al.}~\cite{ksw1,ksw2} as promoted compared to naive 
dimensional analysis. Here the term is further promoted by the identification 
of $\lambda_{\pi{\scriptscriptstyle NN}}$ as a low-energy scale although, in 
practice, this is not crucial since this term cannot be distinguished from 
the one that renormalises the linear divergence. The scale $\mu$ in
this term depends on the choice of renormalisation scheme.

\begin{figure}[h]
\includegraphics[width=8cm,keepaspectratio,clip]{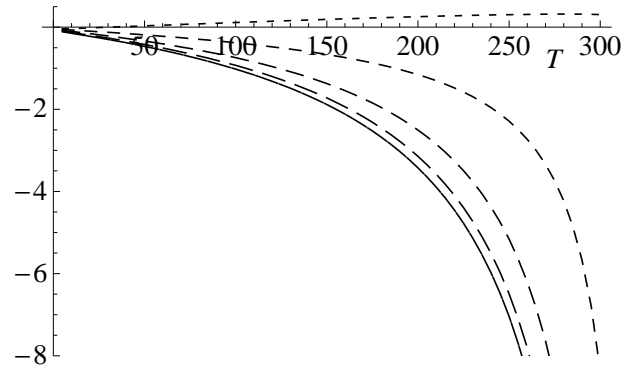}
\caption{\label{fig:v2}  Plots of the inverse of the $^1S_0$ short-distance 
interaction $1/\widetilde V^{(2)}(p)$, in fm$^{-2}$, against lab kinetic 
energy $T$, in MeV. Results are shown for $R_0=1.6$ (shortest dashes), 0.8,
0.4, 0.2 amd 0.1~fm (solid line). The divergent terms in Eq.~(\ref{eq:VS2div})
have been removed by subtracting zero-energy values from all of them.}
\end{figure}

Results for $1/\widetilde V^{(2)}_S(p)$ are shown in Fig.~1. These were 
obtained using the $^1S_0$ phase shift from the Nijmegen partial-wave 
analysis PWA93 \cite{nij93}. Unlike the more peripheral waves studied in 
Refs.~\cite{bmcg,bir07}, the different Nijmegen analyses \cite{nij93,nijpot} 
are in good agreement for this channel and so I present only one here. 
The Nijmegen group's preferred value for the $\pi$N coupling was used, 
$f_{\pi{\scriptscriptstyle NN}}^2=0.075$. Curves are shown for cutoff
radii $R_0$ ranging from 1.6~fm down to 0.1~fm and, to make these easier
to compare, I have subtracted the values at zero energy from each.
This simple-minded renormalisation scheme removes the divergent terms 
identified in Eq.~(\ref{eq:VS2div}) but it means that the momentum scales 
in the leading term of the potential cannot be estimated.

As the radius $R_0$ is taken to zero, artefacts of the cutoff (which are
proportional to positive powers of $R_0$) vanish, leaving a result that 
is independent of $R_0$. The plots show that the $R_0=0.1$~fm has nearly 
converged to its limiting form. In this limit, the expansion of 
$1/\widetilde V^{(2)}_S(p)$ in powers of energy (or $p^2$) is just the DW 
effective-range expansion~\cite{bb1}, and the only effect of the radial 
cutoff $R_0$ is to regulate the linear and logarithmic divergences in the 
energy-independent term. In contrast, for $R_0$ larger than about 1~fm, 
the energy dependence of $V^{(2)}_S$ is dominated by artefacts of the cutoff 
and, by 1.6~fm, and its form is completely different from that of the DW 
effective-range expansion. 

Values of $R_0$ well below 1~fm are obviously outside the expected range 
of validity of this EFT. Nonetheless the power counting used to 
organise the potential remains valid because only OPE is iterated
and so the scaling behaviour of the wave functions is still
controlled by OPE, even in the limit $R_0\rightarrow 0$. This would 
not have been the case if, for example, the TPE potential or 
momentum-dependent contact terms were iterated.

The strong energy-dependence of $\widetilde V^{(2)}_S(p)$ for small $R_0$
reflects the differences between the empirical scattering and that 
produced by OPE alone. The fact that these can be cancelled by artefacts 
of the cutoff (as can be seen most clearly for $R_0=1.6$~fm) is similar 
to a recent observation in Ref.~\cite{bkv} 
for the $^3S_1$--$^3D_1$ channel. The scattering amplitude in that channel 
is known not to converge for momenta of the order of $m_\pi$ \cite{fms} 
when calculated using dimensional regularisation (which affects only the 
potentially divergent pieces of the loop integrals, like the small-$R_0$ 
limit of the approach here). Beane \textit{et al.} found that using an 
additional regulator to soften the core of the tensor potential could lead 
to better convergence. I comment further on this proposal in Sec.~4.

In this context, it is worth noting that for intermediate values of $R_0$, 
around about 0.8~fm, cutoff-dependent effects are still substantial, 
leading to a much smaller, and more weakly energy-dependent, interaction. 
Cutoffs in this regime are often used in applications of EFTs to nuclear 
forces \cite{spm,pv09,em,egm}. For example, the results in Ref.~\cite{pv09} 
were obtained using radial cutoffs in the range 0.6 to 0.9~fm. The present 
results suggest that the finite cutoff is crucial to the good fits to the 
data found there. They may also help to explain why the Nijmegen 
partial-wave analysis \cite{nij93}, which used boundary conditions with 
polynomial dependences on energy, could only give good fits for radii in 
the range $1.4-1.8$~fm.

Finally, the results for small $R_0$ show that, even after removing OPE, 
strong energy dependence remains in the short-range potential for energies 
above about 100 MeV. This matches what Steele and Furnstahl found from 
their application of a DW effective-range expansion to this channel 
\cite{sf}. Since TPE is known to have important effects on the energy 
dependence in this region, this should also be removed before drawing any 
final conclusions about the domain of validity of this EFT.

\section{Two-pion exchange}

Having used DW methods to remove the effects of OPE from the empirical
phase shift, I now turn to TPE. The leading contributions to this
appear at orders $Q^2$ and $Q^3$, and expressions for them can be found
in Refs.~\cite{kbw,nij99}. For consistency, as pointed out by Friar 
\cite{friar}, the leading recoil correction to OPE should also be removed. 
In addition, the leading contribution to $\pi\gamma$ exchange has been 
calculated by Friar \textit{et al.}~\cite{fvkpc} and so this can also be 
dealt with. All of these potentials are perturbations within the power 
counting used here and so the DWBA is sufficient to remove their effects. 
This leaves a potential $V_S^{(4)}$ that retains the effects of long-range
physics only at order $Q^4$ or above.

Starting from the DW $K$ matrix defined in Eq.~(\ref{eq:DWK}), the effects
of the order-$Q^{2,3}$ long-range forces can be removed by the subtracting their 
DWBA matrix elements, taken between the DWs that resum the effects of both OPE 
and the strong short-range potential. This leaves a modified $K$ matrix,
\begin{equation}
\widetilde K_S(p)=\widetilde K(p)-\langle \psi(p)|V_{\rm OPE}^{(2)}
+V_{\rm TPE}^{(2,3)}+V_{\pi\gamma}|\psi(p)\rangle,
\end{equation}
where the DW $\psi(p;r)$ is given by $\psi_>(p;r)$ outside $R_0$ and 
$\psi_R(p;r)$ inside. In order to apply the effective-range method 
outlined above, we need the inverse of $\widetilde K_S$ which, 
also to first order in the long-range perturbations, is given by
\begin{equation}
\frac{1}{\widetilde K_S(p)}=\frac{1}{\widetilde K(p)}
+\left(\!\frac{1}{\widetilde K(p)}\!\right)^{\!2}\!\!
\langle \psi(p)|V_{\rm OPE}^{(2)}+V_{\rm TPE}^{(2,3)}
+V_{\pi\gamma}|\psi(p)\rangle.
\end{equation}
This can then be converted into a residual interaction strength as in
Eq.~(\ref{eq:invVSKS}).

However, before applying the DWBA to the higher-order long-range 
potentials, I need to address the fact that they are highly singular as 
$r\rightarrow 0$ and so their matrix elements between $^1S_0$ waves 
diverge. These must first be made finite,
and a convenient way to do this is to cut off the radial integrals at the 
same radius $R_0$ used to regulate the short-range potential. The divergences 
can then be identified and cancelled by appropriate counterterms, leaving 
only finite quantities to be treated in perturbation theory.

The (unrenormalised) DWBA matrix element is given by the radial integral
\begin{equation}
\langle \psi(p)|V|\psi(p)\rangle
=4\pi\int_{R_0}^\infty r^2\,\mbox{d}r\,V(r)\,\psi_>(p;r)^2.
\end{equation}
In the present case, its strongest divergences arise from the order-$Q^3$ 
piece of TPE, which behaves like $1/r^6$ for small $r$, in combination with 
the irregular parts of the wave functions $\psi_>(p;r)$. The latter behave 
like $1/r$ and so the most divergent term in the radial integral is
\begin{equation}
\int_{R_0}r^2\,\mbox{d}r\,\frac{1}{r^6}\left(\frac{1}{r}\right)^2\propto 
\frac{1}{R_0^5}.
\end{equation}
Higher-order terms in the expansion of $\psi_>(r)$ will contain additional 
powers of $r$ multiplied by low-energy scales. If these scales are $m_\pi$ 
or $\lambda_\pi$, then the corresponding counterterms cannot in practice 
be distinguished from the leading one. Terms containing powers of the 
energy ($p^2$) are therefore of most interest. They arise from terms in 
the expansion of the irregular solution of orders $p^2r$, $p^4r^3$, 
$p^6r^5$, and so on. These lead to divergences proportional to $p^2/R_0^3$ 
and $p^4/R_0$. In contrast all terms of order $p^6$ and above are finite. 
Within the framework of KSW power counting, the required counterterms have 
orders $Q^0$ and $Q^2$. This is consistent with the treatment of long-range 
forces here, which includes terms up to order $Q^3$.

To subtract off the divergent pieces of the integrals, I first use
Eqs.~(\ref{eq:dwI2}, \ref{eq:dwext}) to write the matrix element in terms 
of radial integrals of $V\psi_I^2$, $V\psi_I\psi_R$ and $V\psi_R^2$. I then 
fit the low-energy region ($T=15$--35~MeV) of each of these with a polynomial 
of fourth order in $p^2$, and subtract the divergent terms from 
the integrals.\footnote{More specifically, terms up to order $p^4$ have 
to be subtracted from the integrals of $V_{\rm TPE}^{(2,3)}\psi_I^2$
and $V_{\rm TPE}^{(3)}\psi_I\psi_R$; terms up to order $p^2$ from the
other TPE integrals and from $V_{\pi\gamma}\psi_I^2$; constant terms
from $V_{\pi\gamma}\psi_I\psi_R$ and $V_{\pi\gamma}\psi_R^2$. The 
recoil correction to OPE gives rise to a single divergence, proportional
to $p^2$, in $V_{\rm OPE}^{(2)}\psi_I^2$.} As in the previous section, 
this rather crude renormalisation scheme is adequate for comparing the 
forms of the remaining, nondivergent pieces of the potentials but it does 
not allow an estimate of the scales in the leading terms.

After removing the effects of the order-$Q^{2,3}$ long-range potentials 
from the $K$ matrix using the DWBA, the corresponding short-range 
potential can be determined as before, using
\begin{eqnarray}
\frac{1}{\widetilde V^{(4)}_S(p)}&=&\frac{\psi_R(p;R_0)^2}{\widetilde K_S(p)}\cr
\noalign{\vspace{5pt}}
&&-\,\frac{M_{\scriptscriptstyle N}}{4\pi}\,
\biggl(p\cot\bigl(\delta_R(p)-\delta_I(p)\bigr)\,\psi_R(p;R_0)^2\cr
\noalign{\vspace{5pt}}
&&\qquad\qquad+\frac{\psi_R(p;R_0)\,\psi_I(p;R_0)}{A_R(p) A_I(p)}\biggr).
\end{eqnarray}
The superscript (4) indicates that the only long-range effects that 
remain in this potential are of order $Q^4$ or higher.

\begin{figure}[h]
\includegraphics[width=8cm,keepaspectratio,clip]{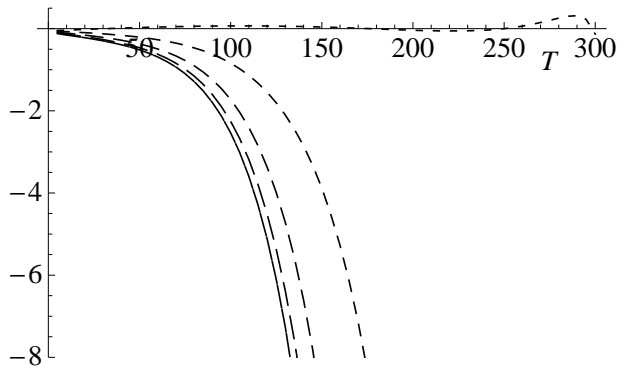}
\caption{\label{fig:v4}  Plots of the inverse of the $^1S_0$ short-distance 
interaction $1/\widetilde V^{(4)}(p)$, in fm$^{-2}$, against lab kinetic 
energy $T$, in MeV. The curves for different $R_0$ are labelled as in Fig.~1.}
\end{figure}

Results for $1/\widetilde V^{(4)}_S(p)$ are shown in Fig.~2. As above,
the empirical input was provided by the Nijmegen 
PWA93 \cite{nij93}. The $\pi$N low-energy constants $c_{1,3,4}$ were taken 
from Ref.~\cite{nij03}, but the results do not change significantly if values 
from other recent determinations are used. As in Fig.~1, the divergent terms in
Eq.~(\ref{eq:VS2div}) have been subtracted to bring the curves for different
radii onto the same plot.

The subtraction of terms up to order $p^2$ from the lar\-gest contributions 
(the integrals of $V_{\rm TPE}^{(2,3)}$) means that the curves are almost 
identical to those in Fig.~1 at very low energies. Effects of TPE start to 
become significant for lab energies above about 70~MeV and then grow 
very rapidly with energy. This rapid increase indicates that any low-energy 
expansion will break down by about $T\sim 150$~MeV, which corresponds to a 
momentum $p\sim 270$~MeV. The only curve that does not show this feature is 
the one for the largest radial cutoff, $R_0=1.6$~fm. However in this case, the
form of the residual potential is controlled by the large artefacts of
the cutoff, like the corresponding curve in Fig.~1. Similar comments to those 
made in Sec.~2 apply to the use of large cutoff radii.

\section{Discussion}

The DW method that was previously applied to NN scattering in peripheral 
waves \cite{bmcg,bir07} has been applied here to the $^1S_0$ channel. The
scattering in this channel is strong at low energies and so the appropriate
EFT is based on a DW effective-range expansion, rather than a Born expansion.
Such an approach was first applied to this channel by Steele and Furnstahl
\cite{sf}, who noted the need to extend it to take account of the effects 
of TPE.

The empirical $^1S_0$ phase shift is deconstructed by first using the DW 
effective-range expansion to remove the effects of OPE to all orders. 
This also takes account of nonperturbative effects of the leading 
(energy-independent) contact interaction, and it generates wave functions 
that are mixtures of the regular and irregular solutions of the 
Schr\"od\-inger equation with the OPE potential. These wave functions are 
then used to evaluate the DWBA matrix elements of the order-$Q^2$ and $Q^3$ 
chiral potentials. Subtracting these leaves a residual short-range 
interaction from which all effects of long-range forces have been removed 
up to order $Q^3$.

Underlying this treatment is the power counting that follows from 
identifying the scale controlling the strength of OPE, 
$\lambda_{\pi{\scriptscriptstyle NN}}$, as a low-energy scale, in addition 
to the momenta, $m_\pi$ and the inverse scattering length $1/a$. In this 
scheme, OPE and the leading contact interaction are of order $Q^{-1}$ and 
hence must be iterated to all orders. All other interactions, including 
TPE, should be treated perturbatively. The $1/r$ singularity of OPE in 
this channel means that, even though that potential is iterated, the power 
counting for the higher-order contact interactions is, for practical 
purposes, the same as that of KSW \cite{ksw1,ksw2}.

The DWBA matrix elements of the TPE potential contain divergent pieces
proportional to $p^0$, $p^2$ and $p^4$, but these can be renormalised using
energy-dependent contact interactions with orders up to $Q^2$ in the 
KSW-like counting. The need for counterterms to this order is consistent
with the fact that long-range forces have been treated to order $Q^3$.
Having used these terms to renormalise the divergences, we can take 
the cutoff radius to be as small as we like (corresponding to an 
arbitrarily large momentum scale). 

This exemplifies the point that, so long as we are careful, it is possible
to take the cutoff beyond the breakdown scale of our EFT. However, as stressed 
in Ref.~\cite{bir09}, this is permissible only if we first determine the 
appropriate power counting, and we then respect that counting. That is, we must 
iterate all terms of order $Q^{-1}$ and we must not iterate any higher-order 
perturbations (``irrelevant'' terms of order $Q^d$ with $d\geq 0$). 
Otherwise, iteration of singular higher-order potentials changes the forms 
of the wave functions at short distances and, in general, destroys any 
consistent power counting.

The present application of the DW method to $^1S_0$ NN scattering has mixed 
implications for EFT descriptions of nuclear forces. The necessary 
counterterms are available to cancel all the divergence terms that arise 
from the DWBA matrix elements of the order-$Q^{2,3}$, as also noted 
in Ref.~\cite{pv09}. The remaining dependence on the cutoff consists of 
artefacts, which vanish like positive powers of $R_0$ as $R_0\rightarrow 0$. 
For $R_0\sim 0.1$~fm, the renormalised residual interaction has essentially
converged to a cutoff-independent form. In contrast, for $R_0$ around 0.8~fm 
or larger, cutoff-dependent effects are substantial and both the size and 
the energy-dependence of the short-range interaction are quite different 
from its $R_0\rightarrow 0$ limit.

Unfortunately the range of applicability of the resulting EFT is limited
since the residual interaction starts to show very strong energy dependence 
in the region $T\gtrsim 100$~MeV. In fact, this energy dependence strengthens 
when the higher-order chiral potentials are subtracted, contrary to the hope 
expressed by Steele and Furnstahl \cite{sf} and also to what might 
have been expected from other channels, where removing long-range forces 
weakens this dependence \cite{bmcg,bir07}. Any attempt to 
expand the residual interaction in powers of the kinetic energy will fail to 
converge for $T\sim 150$~MeV, implying that the breakdown scale is only
of order $\Lambda_0\sim 270$~MeV. 

These results indicate that the EFT developed 
here should not be used in the $^1S_0$ channel for energies above this range. 
They are consistent with the behaviour shown in Fig.~1 of Ref.~\cite{pv09}
which, although not presented in the form of a residual interaction, is based 
on the same method as used here. That shows that, with cutoff radius of 0.1~fm,
a short-range potential consisting of a three-term polynomial in the energy 
cannot provide a good fit to the $^1S_0$ phase shift above about 180~MeV. 
All this implies that the present EFT, which follows the philosophy of Kaplan
\textit{et al.}~\cite{ksw1,ksw2} by treating higher-order terms perturbatively, 
breaks down at a scale of around $2m_\pi$. This is uncomfortably low for 
hopes of applying it to other nuclear systems.

Such a low breakdown scale indicates that there are other low-energy 
scales in this system, in addition to $1/a$, $m_\pi$ and 
$\lambda_{\pi{\scriptscriptstyle NN}}$. An obvious omission from a low-energy 
EFT with only pions and nucleons is the $\Delta$ resonance, which has an 
excitation energy of about 290~MeV. In the present approach effects of the 
$\Delta$ appear only indirectly, via the coefficients $c_{3,4}$ in the 
order-$Q^2$ $\pi$N Lagrangian. The values of these are larger than would be 
natural if the scale of the omitted physics were, say, the mass of the 
$\rho$ meson.

For example, the coefficient of $r^{-6}\exp(-2m_\pi r)$ in the order-$Q^3$ 
TPE potential contains both $\lambda_{\pi{\scriptscriptstyle NN}}$ and 
$c_3\simeq -5$~GeV$^{-1}$. This coefficient, although built out of high-energy 
scales in strict chiral counting, is unnaturally large. This can be seen by 
defining a scale parameter for it by analogy with 
$\lambda_{\pi{\scriptscriptstyle NN}}$ for OPE:
\begin{equation}
\lambda^\prime_{\pi{\scriptscriptstyle NN}}
=\left(\frac{(16\pi)^2 f_\pi^4}{144 g_A^2|c_3|M_{\scriptscriptstyle N}}
\right)^{\!\!\!1/4}\simeq 115\ \mbox{MeV}.
\end{equation}
The small size of this and other scales in the TPE potential is a likely
cause of the low breakdown scale found here. The important, if indirect,
contribution of the $\Delta$ to these scales suggests that one possible 
way to construct a theory with an greater range of validity could be to 
add the $\Delta$ as an explicit degree of freedom, as done in 
Refs.~\cite{orvk,kgw,kem}. It will be interesting to see whether the 
application of the DW method to that EFT can lead to effective interactions 
with less dramatic energy dependences. 

One way forward might be to look for an EFT where at least some of the 
TPE potential could be iterated to all orders. Working with an EFT 
containing the $\Delta$ will promote parts of this potential to lower 
orders but it will still leave them as perturbative effects. Further 
low-energy scales would need to identified if one wanted to promote 
these terms to order $Q^{-1}$ and hence to iterate them in the 
Schr\"odinger equation. These scales would need to be constructed out 
of $F_\pi$ and $M_{\scriptscriptstyle N}$, by analogy with 
$\lambda_{\pi{\scriptscriptstyle NN}}$ in the case of OPE.

A quite different approach to taming this rapid energy dependence could 
be to leave the cutoff finite, even though this is not required 
after perturbative renormalisation of the divergences in the DWBA matrix 
elements. Such treatments have been suggested by various authors
\cite{spkc,dsgs,dgss,bkv,pv09}. It should be stressed the motivation for doing 
this is quite different from the standard one for using a finite cutoff to 
control the unrenormalisable divergences that arise when higher-order 
terms in the potential are iterated by solving the full Schr\"odinger 
equation \cite{em,egm,spm,eg}. Instead one would be using cutoff-dependent 
terms to cancel large but finite contributions that are left after the 
renormalised contributions of the known long-range interactions have 
been removed from the observed scattering.

In Ref.~\cite{bkv}, a similar additional regulator was applied to the 
$^3S_1$--$^3D_1$ channels and this was shown to improve the convergence 
of the original KSW power counting \cite{ksw1,ksw2}, which treats OPE 
perturbatively. Using the same DWBA approach as here, Ref.~\cite{pv09} 
shows that cutoff radii in the range 0.6--0.9~fm can lead to better fits 
to the $^1S_0$ phase shift for energies of 200~MeV and above. 
The results presented here in Figs. 1 and 2 provide some clues as to why 
such methods may work: for cutoff radii around 0.8~fm and above, both
the size and the energy dependence of the residual short-range
interaction are considerably attenuated. As a result the expansion of
this potential in powers of low-energy scales will have a wider range of
convergence. 

From a practical point of view, therefore, such cutoffs can successfully 
eke out the domain of validity of an EFT with a less than ideal separation 
of scales. However, the principles that could justify this remain unclear 
since the use of a cutoff to control contributions that may be large but are 
nonetheless finite is very different from the normal one of regularising
potentially divergent terms in order to renormalise them. 

Beane \textit{et al.}~\cite{bkv} have suggested an analogy with the 
renormalisation scale $\mu$ of perturbative QCD, which is arbitrary but 
can be chosen to optimise the convergence of the resulting expansion. 
Alternatively, it might be viewed as similar to the factorisation 
scale used to define parton distributions in QCD by separating the 
nonperturbative regime from the one where perturbation theory applies. 
Such a role for the radial cutoff of the present approach was in fact 
suggested in Ref.~\cite{bb1}. It remains to be seen whether either of 
these interpretations can be placed on a more rigorous basis, presumably 
with the help of the RG.

\section*{Acknowledgments}

This work was supported by the UK STFC under grants PP/F000448/1 and 
ST/F012047/1, and by the EU Hadron Physics 2 project. Part of it was carried 
out at the Institute for Nuclear Theory, University of Washington, and I am 
grateful to the INT for its hospitality and to the US Department of 
Energy for partial support. I also thank E. Epelbaum, H. Griesshammer, 
D. Kaplan, J. McGovern and, especially, D. Phillips for helpful discussions 
or comments.

\end{document}